\documentclass[12pt,preprint]{aastex}    
\newcommand       \AU           {\,{\rm AU}}

\newcommand	  \g		{\,{\rm g}}
\newcommand       \K            {\,{\rm K}}
\newcommand	  \pc		{\,{\rm pc}}

\newcommand       \mum          {\,{\rm \mu m}}
\newcommand	  \Teff	        {T_{\rm eff}}
\newcommand	  \amin	        {a_{\rm min}}
\newcommand	  \amax	        {a_{\rm max}}
\newcommand	  \rmin	        {r_{\rm min}}
\newcommand	  \rmax	        {r_{\rm max}}

\newcommand	  \rp           {r_{\rm p}}

\newcommand       \Lsun         {L_\odot}

\newcommand       \Lstar        {L_\star}
\newcommand       \simali       {\sim\,}

\input{psfig.sty}

\shorttitle{RV Boo}
\shortauthors{Biller et al.}

\begin{document}           

\title{High Resolution Mid - Infrared Imaging of the AGB Star 
RV Boo with the Steward Observatory Adaptive Optics System}

\author{B.A. Biller\altaffilmark{1}, L.M. Close, A. Li\altaffilmark{2}, 
J.H. Bieging, W.F. Hoffmann, 
P.M. Hinz, D. Miller, G. Brusa, M. Lloyd-Hart, F. Wildi, D. Potter, 
B.D. Oppenheimer}
\affil{Steward Observatory, University of Arizona, Tucson, AZ 85721}

\altaffiltext{1}{email: \sf{bbiller@as.arizona.edu}}
\altaffiltext{2}{Theoretical Astrophysics Program,
        Steward Observatory and
        Lunar and Planetary Laboratory,
        University of Arizona, Tucson, AZ 85721; email:
        {\sf agli@lpl.arizona.edu}}

\date{}                     

\begin{abstract}
We present high resolution ($\sim$0.1$\arcsec$), very high Strehl ratio 
(0.97$\pm$0.03) 
mid-infrared (IR) adaptive optics (AO) images of the AGB star 
RV Boo utilizing the MMT adaptive secondary AO system.  
RV Boo was observed at a number of wavelengths over two epochs 
(9.8 $\mu$m in May 2003, 8.8, 9.8 and 11.7 $\mu$m in February 2004)
and appeared slightly extended at all wavelengths.  While the extension is 
very slight at 8.8 and 11.7 $\mu$m data, 
the extension is somewhat more pronounced at 9.8 $\mu$m. 
With such high Strehls we can achieve
super-resolutions of 0.1$\arcsec$ by deconvolving RV Boo with a point-spread
function (PSF) derived 
from an unresolved star.
We tentatively resolve RV Boo into a 0.16$\arcsec$ FWHM
extension at a position angle of 120 degrees.  At a distance of 
390$^{+250}_{-100}$ pc, 
this corresponds to a FWHM of 60$^{+40}_{-15}$ AU.  
We measure a total flux
at 9.8 $\mu$m of 145$\pm$24 Jy for the disk and star.  Based on a dust thermal 
emission model for the observed IR spectral energy distribution and the 
9.8 $\mu$m AO image, 
we derive a disk dust mass of 1.6$\times$10$^{-6}$ M$_{\sun}$ 
and an inclination of 30 to 45$^{\circ}$ from edge-on.  
We discuss whether the dust disk observed around RV Boo
is an example of the early stages in the formation of asymmetric 
structure in planetary nebula.
\end{abstract}


\section{Introduction \label{sec:intro}}

Extensive mass loss in the asymptotic giant branch (AGB) 
phase has been well established.  
However, the mode (or modes) by  
which this mass loss occurs is less well known.  While some objects 
present CO lines indicative of spherically symmetric mass loss, other objects 
display much more complicated mass loss line profiles 
\citep{multwinds, COsurvey}.  A small number of 
AGB and post-AGB stars display a
particular type of anomalous CO features: a 
very narrow peak (as narrow as $\sim$1 km s$^{-1}$, 
see Kahane et al. 1998, but 
generally $\leq$5 km s$^{-1}$), with 
or without a broader underlying pedestal feature 
(with widths of 10-20 km s$^{-1}$).  
Stars which display just the narrow peak include AC Her, BM Gem, and the 
Red Rectangle.  X Her \citep{XHer}, RV Boo (Bergman, Kerschbaum, \& Oloffson 
2000), EP Aqr, RS Cnc, and
IRC +50049 display narrow peaks as well as underlying broader components. 
\citet{MolRes} interpret the narrow features in these objects as reservoirs 
of 
dust and molecular gas which are nearly at rest with 
respect to the central AGB stars.  
They suggest that a binary companion is necessary in these 
cases in order to entrain gas and 
dust into a circumbinary disk \citep{binary, Mas1, Mas2}; 
both AC Her and the Red Rectangle
have companions.  
The broader CO line components are interpreted as spherical outflows
or in some cases (RS Cnc and X Her) as bipolar outflows \citep{XHer}.  
RV Boo is unusual since interferometric images of its CO emission suggest 
the presence of a large disk in Keplerian rotation \citep{RVBoo1}.  
Interestingly, \citet{DV} has also recently imaged another 
sort of asymmetry in AGB star envelopes -- a small
bipolar outflow observed in the near-IR around the AGB star IRC+10011. 

Determining the spatial structure of winds around AGB stars is important 
for constraining models of bipolar planetary nebulae.
Generalized Interacting Stellar Winds (GISW) models of 
planetary nebulae invoke some initial structure 
which can collimate and shape the fast
winds produced by these objects.  In these models, the existence or 
non-existence of this structure regulates whether the forming planetary nebula
acquires a round, elliptical, or bipolar morphology \citep{PNrev}.  
This pre-existing structure is likely to have 
formed by the end of the central object's AGB stage, since a sizable percentage
of proto-planetary nebulae display bipolar reflection nebulae.
Could the disks and molecular
reservoirs observed in CO around AGB stars be the initial stages
in the formation of such pre-existing
structure?  The molecular reservoirs observed by \citet{BMGem1}
may be the diffuse precursors to the formation of a denser disk
or torus around the star at the end of the AGB phase.
Such dense disks might be capable of collimating the fast winds produced
when these objects evolve to the PN phase.  

At mid-IR wavelengths, we expect dusty disks around AGB and post-AGB
stars to be detected in thermal emission.  
Do we observe the disks implied by CO
observations at these wavelengths?  \citet{PPNsurvey} observed a number
of protoplanetary nebulae (pPNe) in the mid-IR.  
In those which were resolved, they
found two primary mid-IR morphologies -- core/elliptical and toroidal.
Ueta, Meixner, \& Bobrowsky (2000) found that each of these 
mid-IR morphologies also corresponds to a 
specific optical morphology.  The optically thick
core/elliptical mid-IR morphologies possessed bipolar reflection nebulae
and heavily obscured central stars.  The optically thin toroidal mid-IR
morphologies possessed elliptical reflection nebulae and non-obscured
central stars.  In a number of cases dusty disks have not 
been observed directly but
may collimate reflection nebulae -- for instance, \citet{RedRect2}
observe double lobed reflection in the NIR (J and K bands)
around the Red Rectangle (IRAS 06176-1036) and Frosty Leo (IRAS 09371+1212).
\citet{PPNsurvey} resolved the Red
Rectangle into an unresolved core and an extended elliptical nebulosity but
found Frosty Leo to be unresolved.

RV Boo is an O-rich type b semiregular variable.  It varies in luminosity and
in spectral class between M5III and M7III with a period of $\sim$140 days.  
Previously, \citet{RVBoo1} observed a 4$\arcsec$ 
diameter disk in CO around RV Boo.  They interpret this disk as 
possibly the first known Keplerian disk around an 
AGB star.  The resolution in CO radio lines is limited to $\sim$1-2$\arcsec$ 
even using interferometry; by observing at high resolution in the mid-IR, 
we can probe the structure of this disk on much finer scales.

Using the unique adaptive secondary mirror 
AO system at the 6.5m MMT (Wildi et al. 2003, Brusa et al. 2003), we can 
observe AGB stars at mid-IR wavelengths with $\sim$0.1\arcsec resolution
\citep{ACHer2}.  Through deconvolution, the nearly perfect images 
(Strehl ratio$\sim$0.97$\pm$0.03) produced with AO at the MMT 
allow resolutions better than that of the 
FWHM of the diffraction limit of the telescope. 
With such resolutions, we can probe AGB star and PPN morphologies on finer 
scales than ever before possible in the mid-IR.  Here, we 
present the first adaptive optics high resolution images of RV Boo.  

\section{Observations and Data Reduction \label{sec:obs}}

Data were taken on the night of 2003 May 13 (UT) at the 6.5m MMT using 
the adaptive secondary mirror AO system with the 
BLINC-MIRAC3 camera \citep{MIRAC, MIRAC2}.  
The adaptive secondary corrected the first 
52 system modes at 550 Hz and achieved 
Strehl ratios as high as 0.97$\pm0.03$ from 8.8 - 18 $\mu$m.  These Strehl
ratios are the highest ever presented in the literature; previous  
Strehl ratios for large telescopes have rarely exceeded 0.7 at any 
wavelength.  Our unique ability to do AO correction at 10 $\mu$m leads to 
very high Strehl ratios regardless of the seeing, airmass, or wind 
\citep{ACHer2}.  With the very stable PSFs 
that result from such high Strehls it is possible to detect structures with 
spatial scales smaller than the diffraction-limited FWHM 
($\sim$0.98$\frac{\lambda}{D}$ rad) through the use of deconvolution.

\begin{figure}
\plotone{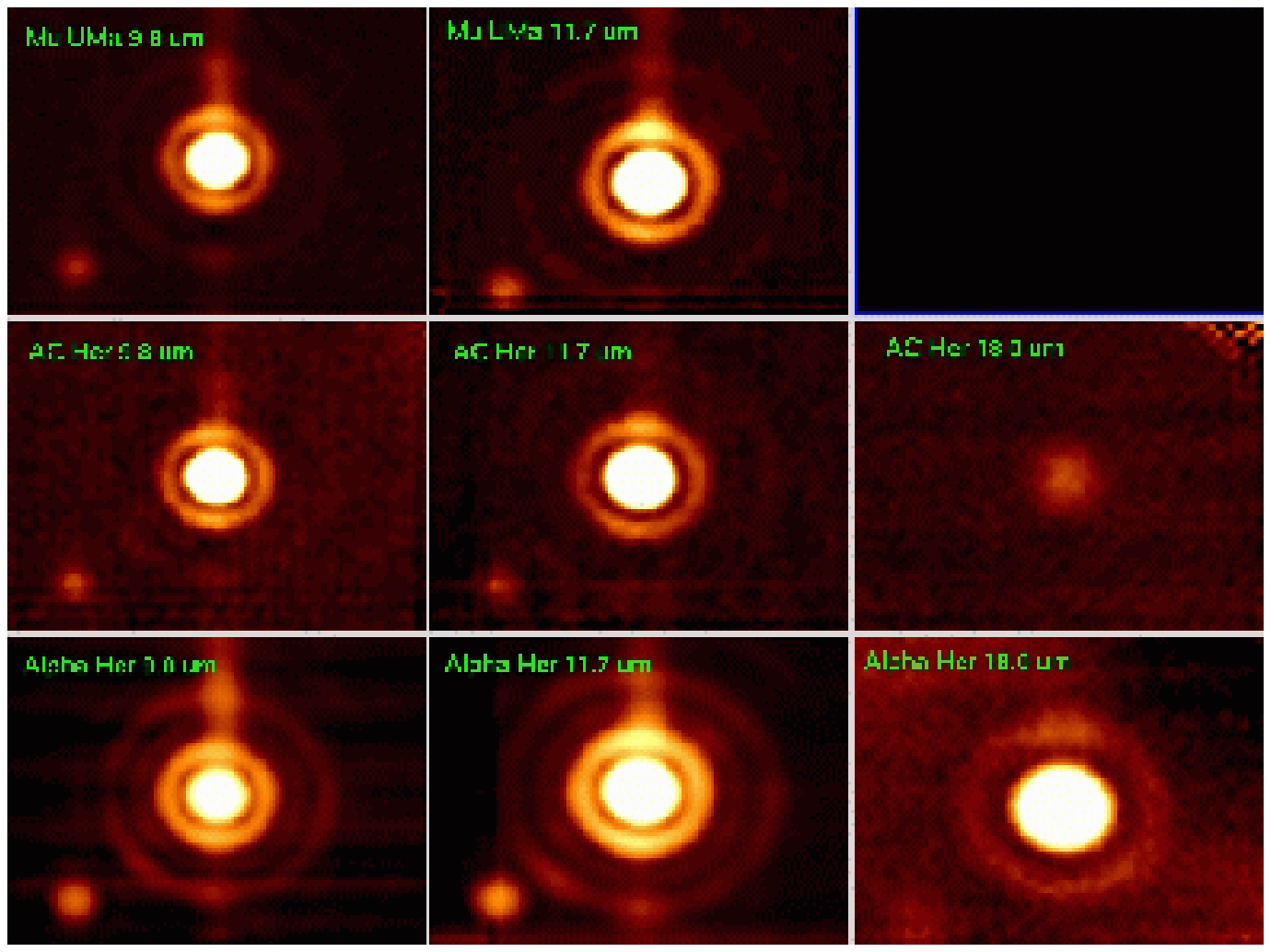}
\caption{The 9.8, 11.7, and 18 $\mu$m images of the 
PSF stars AC Her, $\mu$ UMa, 
and $\alpha$ Her as observed at the MMT.  The box size of the MMT images is 
$1.5\times1.0\arcsec$.  The faint point source in the lower left of each 
MMT image is a MIRAC3 ghost.  (Fig.~1 of \citet{ACHer2}.)}
\label{fig:PSFS}
\end{figure}

\begin{figure}
\plotone{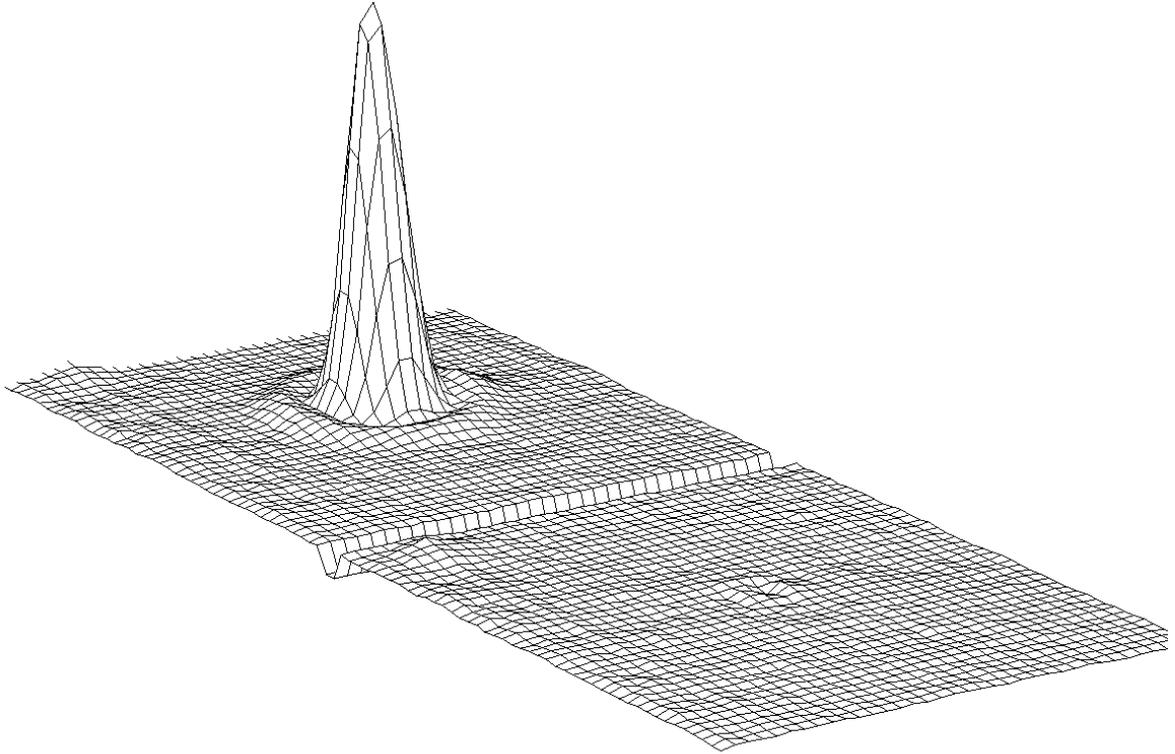}
\caption{The 11.7 $\mu$m PSF of AC Her before (left) and after (right) PSF
subtraction (using $\alpha$ Her as the PSF) with DAOPHOT's ALLSTAR task.  The
residual flux after PSF subtraction is $<0.5\%$ of AC Her's original flux.
Similar residuals resulted from PSF subtractions at 9.8 $\mu$m and 18 $\mu$m.
Based on these excellent subtractions, we conclude that 
the PSF obtained from the MIRAC3 camera with the MMT adaptive secondary 
AO system is extremely stable.  Note that the small ghost image to 
the lower left in each frame is not subtracted to show that the vertical 
scales are the same for both images.  (Fig.~3 of \citet{ACHer2}.)}
\label{fig:PSFsubtraction}
\end{figure}

Images of all the PSF stars observed on 2003 May 13 are displayed in 
Fig.~\ref{fig:PSFS}.  To further illustrate 
stability, we subtracted one PSF star
($\alpha$ Her) from another observed later in the evening (AC Her).  The 
residuals are displayed in Fig.~\ref{fig:PSFsubtraction}.  The
residual flux after PSF subtraction is $<0.5\%$ of AC Her's original flux.
Similar residuals resulted from PSF subtractions at 9.8 $\mu$m and 18 $\mu$m.
Based on these excellent subtractions, we conclude that 
the PSF obtained from the MIRAC3 camera with the MMT adaptive secondary 
AO system is extremely stable.   

RV Boo was observed in the 9.8 $\mu$m wavelength band.  Point sources 
$\mu$ UMa and $\alpha$ Her were observed in this band
before and after RV Boo to use as PSF calibrators.  $\alpha$ Her is a 
relatively wide binary with a separation of 4.7$\arcsec$ \citep{binarymeas}; 
the  brighter component was used as a PSF while the fainter 
component falls outside
our field of view.  We note that 
$\alpha$ Her also does not possess a dust shell or 
other extended structure \citep{ACHer2}.
To eliminate the high sky background at mid-IR 
wavelengths, we used a standard chopping/nodding scheme and flat-fielded our 
data.  A chopping frequency of 1 Hz (throw$\sim$20$\arcsec$) was 
used with a nodding cycle of 60 sec (throw$\sim$6-8$\arcsec$).  
To avoid saturation of the high sky background, a base integration 
time of 29 msec was used.  These images were coadded to produce an output
frame every 15 seconds.
For RV Boo, 4 15-second integrations were 
taken at each of 8 nod positions, giving a total exposure time of 8 minutes.  
$\mu$ UMa was observed using 8 nod positions, for a 
total exposure time of 8 minutes.  
$\alpha$ Her was observed using 4 nod positions, for a total exposure time of 
4 minutes.  We used the internal BLINC cold chopper
and kept the AO in closed loop for both chop and nod beam positions.   

Flat fields were taken the night of 2003 May 15 (UT).  
The base integration time was set to 10 msec.  
These images were coadded every 2 seconds.  
Flats were taken of the inside of the dome (hot) and 
the sky (cold).  
The sky flats were subtracted from the dome flats.  The resulting
flat fields were normalized by the mean.

To determine an astrometric calibration, we used our 25 November 2002 (UT)
observations of the binary star WDS 02589+2137 BU.  These data were taken in
the M band using MIRAC with the MMT adaptive secondary AO system.  At
the time of observation, the binary had a position angle of 269$^{\circ}$ and 
a separation of 0.509$\arcsec$ \citep{WDS6thcat}.  
To align our RV Boo data with North, we must rotate it by 
(270$^{\circ}$ -- the parallactic angle) at the time of 
observation.  We also determine a plate scale of 
88 milliarcsec/pixel from this standard.
  
\begin{figure}
\plotone{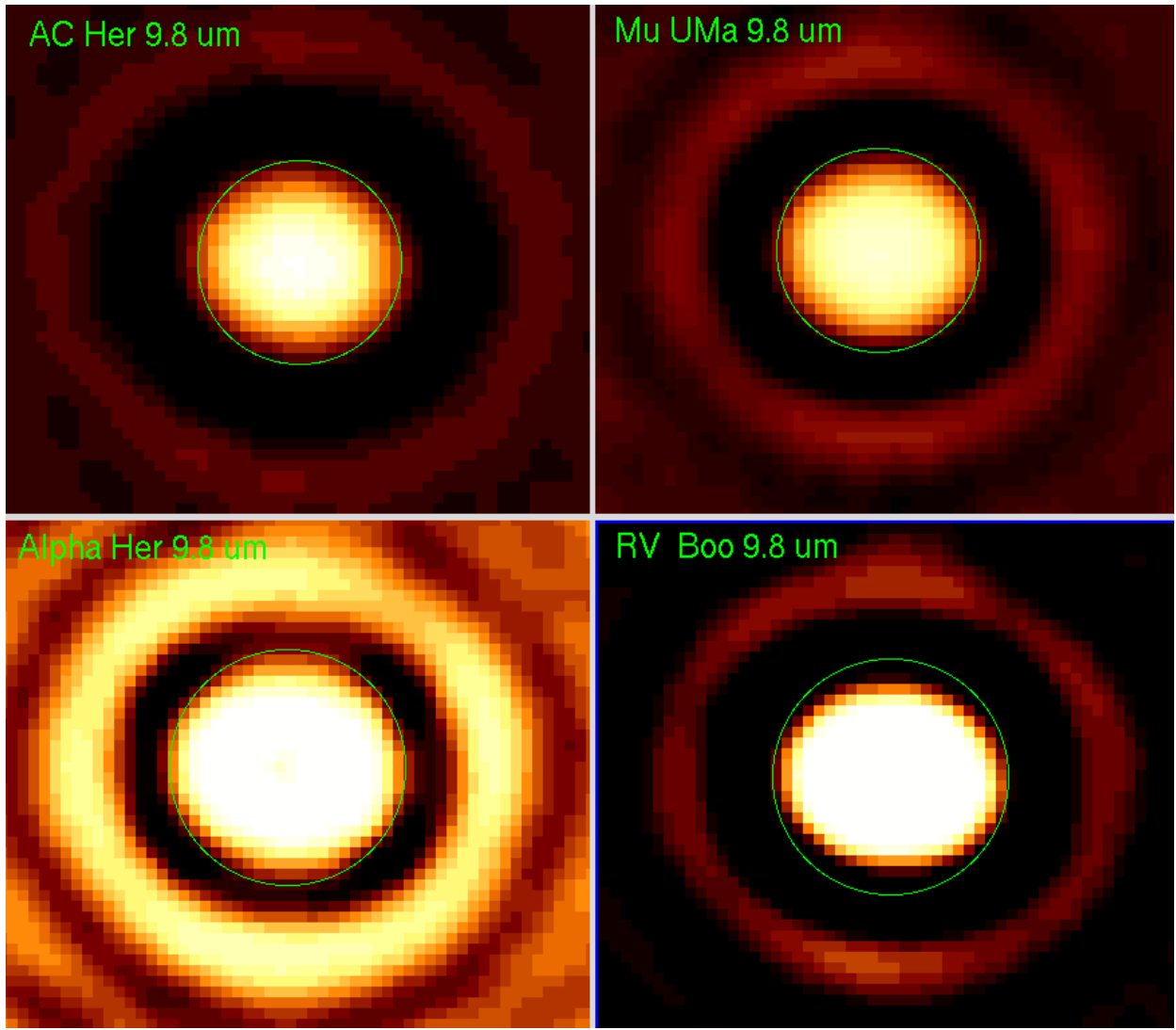}
\caption{AO Images of RV Boo, $\mu$ UMa, $\alpha$ Her, and 
AC Her at 9.8 $\mu$m.  The vertical axis is telescope altitude while the 
horizontal axis is telescope azimuth.  The images are Fourier filtered: 
An image smoothed with a Gaussian with a 3 pixel FWHM 
 was subtracted from each of the 
original images in order to remove variations on large spatial wavelengths.
All images are shown on a logarithmic scale; the bright
ring around the images is the first Airy ring.  Note that RV Boo appears 
nominally extended relative to the other stars.  
The PSF star $\alpha$ Her is slightly 
saturated. A 0.7$\arcsec$ diameter 
 circle is overlaid on each image to help aid the eye.}
\label{fig:RVBoo_raw}
\end{figure}

Data were reduced using a custom IRAF pipeline which first 
flat fields and removes bad pixels.  After the pipeline completes these 
basic data reduction tasks, it then rotates the nod images by 
(270$^{\circ}$-- the parallactic angle) 
and coadds them so that North is up and East is left.  
A coadded image of RV Boo alongside similar
images of the PSF stars $\mu$ Uma and $\alpha$ Her as well as the AGB star
AC Her \citep{ACHer2} is presented in Fig.~\ref{fig:RVBoo_raw}.  
The vertical axis is telescope altitude while the horizontal axis is
telescope azimuth.  RV Boo appears slightly extended (FWHM$\sim$4 pixels)
relative to the PSF stars (FWHM$\sim$3.8 pixels.)
All three PSF stars are very slightly (eccentricity$\sim$2$\%$) 
elliptical along the horizontal (azimuthal) direction 
-- this is a systematic instrumental feature of the PSF.  
However, RV Boo appears somewhat more elliptical (eccentricity$\sim$5$\%$) 
than the other three stars.
We argue that this extension and slight ellipticity is indicative of actual 
physical structure.  The observed extension lies near the limit of 
resolution of the telescope, however, and is broadened by
diffraction.  To discern the actual small 
scale structure of the extension around RV Boo, 
we must deconvolve it with a PSF star. 

In order to determine if the extension we see is real (and not just the 
result of a vibration in the telescope mount, for instance), we 
deconvolved each of the 8 RV Boo nod images with the 
$\mu$ UMa PSF.  After subpixel interpolation by 3x (to a new platescale 
of 29.3 milliarcsec/pixel), we used the Lucy deconvolution algorithm
in IRAF with 1000 iterations.  We chose to deconvolve for 1000 iterations for
two reasons -- first, 
object properties (position angle and deconvolution) vary 
rapidly up until the $\sim$800th iteration.  By the 1000th iteration,
properties have converged.  Secondly,  
we convolved a set of thermal disk models 
at inclination angles from edge on of 5, 15, and 30$^{\circ}$ (see 
\S\ref{sec:analysis} for details on our modeling) with the $\mu$ UMa PSF, then
deconvolved for 100, 500, 1000, 1500, and 2000 iterations
using AC Her \citep{ACHer2} as the PSF.  The 1000 iteration deconvolutions 
best recreated the position angles and eccentricities of our 
models.  

If the extension is real 
we expect the position angle of the deconvolved semi-major axis to track the 
parallactic angle as the sky rotates between exposures.  
We measured position angle
and eccentricity for each deconvolution 
of the 8 individual RV Boo nod images using the $\it{imexam}$ tool
in IRAF.  The $\it{imexam}$ tool is only accurate for measuring isophote
semi-major position angles to within $\pm$6$^{\circ}$ of accuracy.  
We calculate $\sigma$=5.65$^{\circ}$ as our 
error in position angle measurements.  This error was determined by
stretching an image of $\mu$ UMa to 5$\%$ eccentricity (about the eccentricity
of RV Boo previous to deconvolution), rotating it through a number of 
angles, and then taking the standard deviation of (measured angle -
actual angle) as the error.

\begin{figure}
\plotone{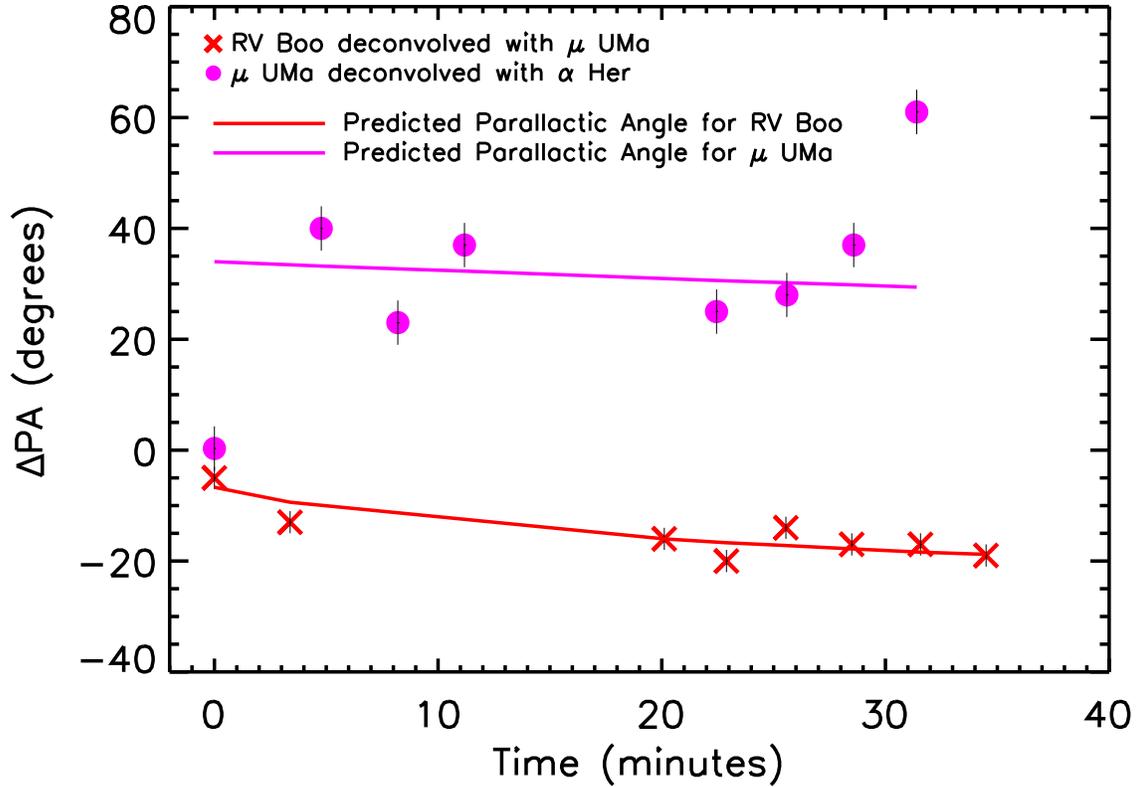}
\caption{Position angle of the semi-major axis vs. 
time (after first observation) for deconvolved RV Boo and $\mu$ UMa 
nod images.  Measured position angles are represented as crosses for RV Boo
and circles for $\mu$ UMa; solid lines depict the predicted parallactic
angle as a function of time during the observations. 
Note that the $\Delta$PAs measured for RV Boo track the
parallactic angle much more closely than those measured for $\mu$ UMa, 
with a reduced $\chi^2$ value of 0.207 for RV Boo deconvolved with
$\mu$ UMa versus a reduced $\chi^2$ value of 
13.2 for $\mu$ UMa deconvolved with $\alpha$ Her.  This implies that the
elongation observed was really associated with RV Boo (since it was 
rotating along with the sky) and is not a PSF artifact.
}
\label{fig:PA}
\end{figure}

We plot the behavior of the major axis position
angle of the deconvolved image (henceforth $\Delta$PA)  
and parallactic angle with time in Figure~\ref{fig:PA}.   
We fit the expected parallactic angles (calculated using
the SKYCALC software package -- \citet{skycalc}) to the position angle data 
for each star with a minimized $\chi^2$. 
Thus, $\Delta$PA measures how
much the trend in position angle deviates from the trend in parallactic angle.
Expected parallactic angles are plotted 
as solid lines; observed values of $\Delta$PA are plotted using a variety 
of points.   We plot the behavior of 
$\Delta$PA for RV Boo deconvolved with $\mu$ UMa as the PSF.  
As a comparison, we plot 
$\Delta$PA for $\mu$ UMa deconvolved
with $\alpha$ Her, a PSF deconvolving a PSF (both apparent point
sources).  The $\Delta$PA values measured from the 
RV Boo deconvolutions follow 
the correct rotation of the sky with a reduced $\chi^2$ of 0.207 for the 
deconvolution with $\mu$ UMa.  Thus, RV Boo's extension is rotating on the
sky to within a probability of 98$\%$. 
In contrast, the $\Delta$PA values measured from the 
PSF and PSF deconvolution show much more scatter and have a best fit to   
the parallactic angles with a reduced $\chi^2$ of 13.2.  Hence, the extension
of a PSF deconvolved with a PSF is purely an artifact; the probability that 
$\Delta$PA for $\mu$ UMa deconvolved with $\alpha$ Her rotates on the sky
is less than 0.1$\%$ as one would expect.  

\begin{figure}
\plotone{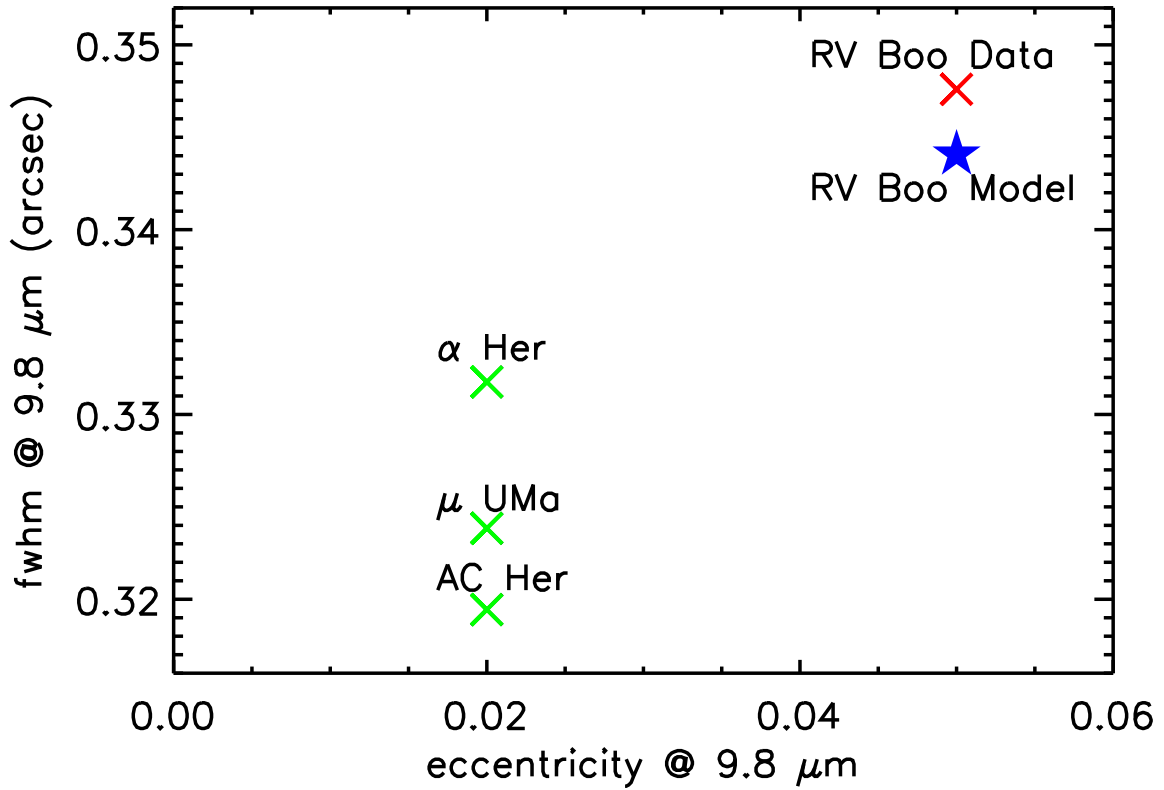}
\caption{Eccentricity vs. PSF FWHM for RV Boo, $\mu$ UMa, 
$\alpha$ Her, and AC Her images.  The best fit thermal emission model
 of RV Boo convolved with the $\mu$ UMa PSF is also plotted 
(see \S\ref{sec:analysis}
for details on modeling).  The three PSF stars are plotted as green crosses.
The RV Boo data is plotted as a red cross and the RV Boo model is plotted 
as a blue star.  FWHM is measured by a Gaussian fit to the 
enclosed flux at each radius. 
RV Boo appears slightly extended and 
has a significantly higher eccentricity and FWHM than the other stars.}
\label{fig:E}
\end{figure}

The variation of PSF FWHM with source eccentricity for
RV Boo, $\mu$ UMa, $\alpha$ Her, and AC Her is presented in Fig.~\ref{fig:E}.  
These quantities are also presented in Fig.~\ref{fig:E} 
for the best fit thermal emission model of the RV Boo disk 
convolved with the $\mu$ UMa PSF
(see \S\ref{sec:analysis}).  An extended source should have
a larger FWHM and possibly a higher eccentricity (depending on position 
angle, source shape, etc.) than a point source; RV Boo indeed
shows this trend.  
The RV Boo disk models accurately recreate the FWHM and eccentricity of
the RV Boo data.  Models were fit to the deconvolved data;
the match between convolved model and undeconvolved data  
implies that 1000 iterations of the Lucy algorithm produces an 
accurate deconvolution of the RV Boo data.  
Based on the rotation of the extension with the sky 
and its non-PSF FWHM and eccentricity, 
we tentatively conclude that the extension 
we observe around RV Boo
is indeed real and not an artifact of the telescope.  However, to
strengthen this conclusion, we reobserved RV Boo
at a variety of wavelengths at another epoch.
 
\section{Followup Observations \label{sec:followup}}

RV Boo was reobserved on the night of 2004 February 2 (UT) 
at the 6.5m MMT using 
the adaptive secondary mirror AO system with the 
BLINC-MIRAC3 camera \citep{MIRAC, MIRAC2}. Similar Strehl Ratios
were observed as during the first observations, however, the MIRAC camera 
was used with a platescale of 0.079$\arcsec$ (as opposed to 0.088$\arcsec$
during the previous run.)  These data was taken during an engineering
test and are of considerably lower quality (engineering grade 
rather than science grade) than the May 2003 dataset. 

As before, we used a standard chopping/nodding scheme.  
A chopping frequency of 1 Hz (throw$\sim$20$\arcsec$) was 
used.  Each object was observed at three different wavelengths (8.8 $\mu$m,
9.8 $\mu$m and 11.7 $\mu$m) and two different nod positions 
(throw$\sim$4-5$\arcsec$).
To avoid saturation of the high sky background, a base integration 
time of 30-70 msec was used.  These images were coadded to produce an output
frame every 10 seconds.
For RV Boo, 10x10s integrations were 
taken at 2 nod positions for each wavelength, 
giving a total exposure time of 200 seconds per wavelength.  
$\alpha$ Boo was observed at 2 nod positions per wavelength, for a 
total exposure time of 200 seconds per wavelength.  
We used the internal BLINC cold chopper
and kept the AO in closed loop for all chop and nod beam positions.  
The data were reduced with the same custom IRAF pipeline as before.
With shorter total integration times (200 s per wavelength 
for RV Boo during reobservations vs. 
480 s at 9.8 $\mu$m during the initial observations), 
our signal to noise during the followup observations is only $\sim$65$\%$
that achieved during the initial observations.

PSF and data FWHMs for both the May 2003 and 
February 2004 datasets are presented in Table~\ref{tab:FWHM}.
We estimate an uncertainty in our measurements of $\pm$6 milliarcsec 
from the scatter of the PSF FWHMs in the May data.
The FWHM of RV Boo was slightly larger than that of the PSF 
in all wavelengths in February, however, the extension is within the 
measurement error for the 8.8 and 11.7 $\mu$m data.  The February RV Boo 
9.8 $\mu$m data possesses a FWHM comparable to that of the May 
RV Boo data -- thus, we reobserve the slight extension found in May in 
our February dataset.  Perhaps it is not surprising that the object appears
most extended at 9.8 $\mu$m -- right at the center of a prominent 
silicon emission feature seen in the ISO spectrum (see $\S$\ref{sec:analysis}
and Fig.~\ref{fig:RVBoo_sed}).  For RV Boo, we measured eccentricities of 
$\sim$2$\%$ in February vs. $\sim$5$\%$ in May.   
This may have simply been a function of 
position angle during the observations.  Unfortunately, 
the engineering grade observations of the only PSF 
star ($\alpha$ Boo) were found to have low spatial frequency noise and were 
unsuitable to be used as a PSF for deconvolution.  As a test, we deconvolved
these observations using the $\mu$ UMa PSF from the previous observations.  
The February data possessed similar deconvolved source properties 
(FWHM, position angle, etc.) compared to the May data.

\begin{deluxetable}{cccccccc}
\tablecolumns{8}
\tablecaption{FWHMs for RV Boo and PSF stars \label{tab:FWHM}}
\tabletypesize{\small}
\tablewidth{6.5in}
\tablehead{
\colhead{} & \multicolumn{4}{c}{May 2003} & \colhead{} & \multicolumn{2}{c}{February 2004}\\
\cline{2-5} \cline{7-8} \\
\colhead{$\lambda$} & \colhead{$\mu$ UMa (PSF)$^*$} & \colhead{$\alpha$ Her (PSF)} & \colhead{AC Her (PSF)} & \colhead{RV Boo} & \colhead{} & \colhead{$\alpha$ Boo (PSF)} & \colhead{RV Boo}\\}
\startdata
8.8 $\mu$m & \nodata & \nodata & \nodata & \nodata &
& 0.292$\arcsec$ & 0.297$\arcsec$ \\
9.8 $\mu$m & 0.324$\arcsec$ & 0.332$\arcsec$ & 
0.319$\arcsec$ & 0.348$\arcsec$ &
& 0.322$\arcsec$ & 0.338$\arcsec$ \\ 
11.7 $\mu$m & \nodata & \nodata & \nodata & \nodata &
& 0.382$\arcsec$ & 0.393$\arcsec$ \\
\enddata

~$^*$The uncertainty for all of the FWHMs is $\pm$6 milliarcsec.
\end{deluxetable}

\section{Analysis \label{sec:analysis}}

\begin{figure}
\plotone{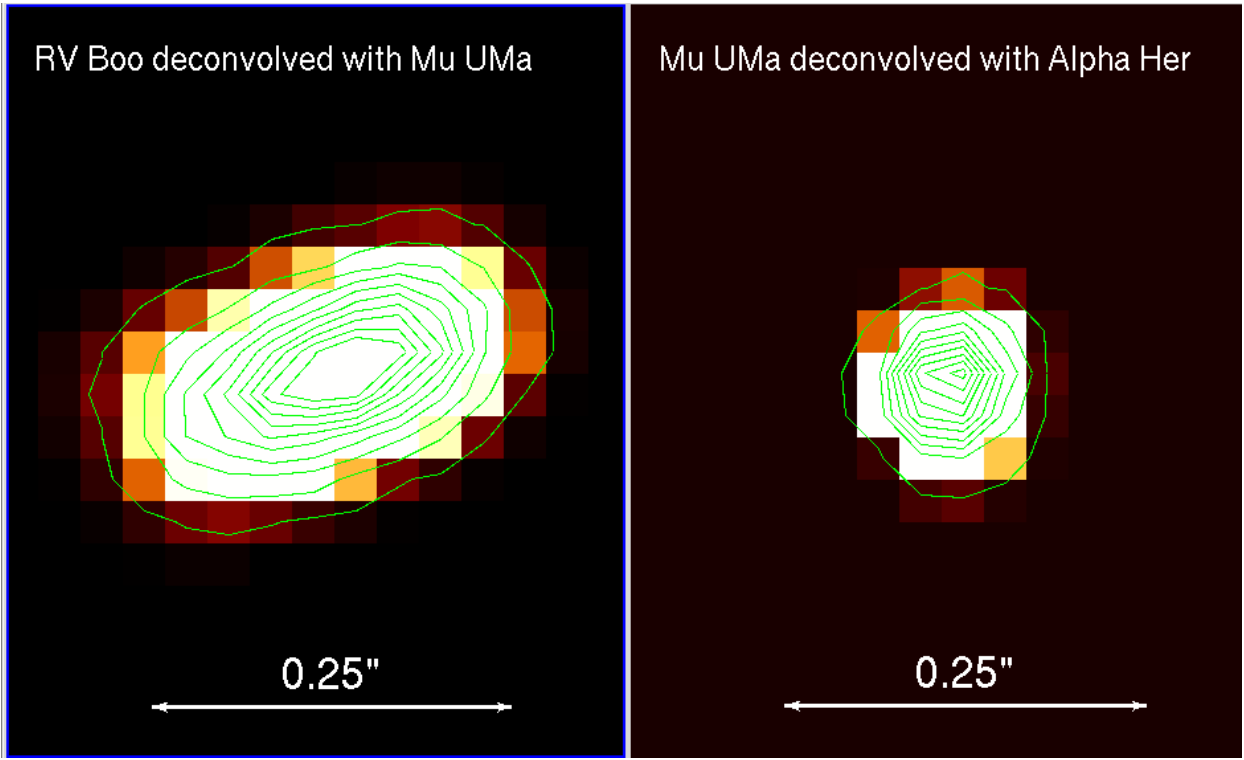}
\caption{Deconvolved image of RV Boo.  The platescale (after magnification)
is 29.3 milliarcsec/pixel.  
North is up, east is left.  The disk is at a position 
angle of 120$^{\circ}$ and has a major axis FWHM of 0.16$\arcsec$ 
($\sim$60 AU at 390 pc).  A deconvolved image of $\mu$ UMa (deconvolved using
$\alpha$ Her as a PSF) is shown on the right for comparison.
For each object, there are 10 contours spaced linearly.  
For RV Boo, the lowest contour level is placed at 3$\%$ of the peak flux
(275 Jy arcsec$^{-2}$) and the highest
contour level is placed at 98$\%$ of the peak flux (9640 Jy arcsec$^{-2}$).
For $\mu$ UMa, the lowest contour level is placed at 1$\%$ 
of the peak flux (279 Jy arcsec$^{-2}$) 
and the highest contour level is placed at 95$\%$ of the peak 
flux (2.24$\times$10$^4$ Jy arcsec$^{-2}$).  After deconvolution, RV Boo
appears extended, while the PSF still appears pointlike.}
\label{fig:RVBoo_deconv}
\end{figure}

The deconvolved image of RV Boo is presented in Fig.~\ref{fig:RVBoo_deconv}.  
This image was produced by coadding all of the RV Boo nod images which had
been deconvolved for 1000 iterations using the Lucy algorithm 
with $\mu$ UMa as the PSF.  (Due to the slight saturation of the 
$\alpha$ Her image, $\mu$ UMa was chosen as the 
PSF over $\alpha$ Her for this analysis.)
After deconvolution, the extended structure
appears to be a disk seen at an inclination angle between 20 to 30$^{\circ}$ 
from edge-on (inclination estimated from modeling; see below).  
From their CO J=2-1 interferometric data, 
\citet{RVBoo1} found evidence of a rotating disk around RV Boo in the
form of a $\sim$4$\arcsec$ diameter disk with a position angle 
of $\sim$150$^{\circ}$.
Using the $\it{imexam}$ tool in IRAF, we measured a FWHM at 
$\lambda$=9.8 $\mu$m of 0.16$\arcsec$ along the major axis and a PA of 
120$^{\circ}$.     
The 9.8 $\mu$m emission traces dust thermal re-emission; RV Boo heats only a 
fraction of the dust in the entire CO disk to temperatures where dust 
is luminous in the mid-IR.
If circumstellar gas and dust are well-mixed around RV Boo, we would expect 
to find a much smaller disk extent at 9.8 $\mu$m than in CO, since all of the
CO will emit, but only the dust near enough to the star to be heated to 
temperatures of $\sim$300 K will emit at 9.8 $\mu$m.
To map out more extension in the dust disk, 
observations at longer wavelengths (which 
trace a larger region of dust re-emission) are necessary.  
We measure a total flux for RV Boo at 9.8 $\mu$m of 145$\pm$24 Jy.

At a distance of 390$^{+250}_{-100}$ pc \citep{hipref}, 
the mid-IR disk of RV Boo subtends
a FWHM of 60$^{+40}_{-15}$ AU.  RV Boo
is a relatively nearby AGB star; it would be more difficult 
to observe a similar disk around a more distant star like AC Her which is 
$\sim$2 times more distant.
Since RV Boo is so nearby relative to the distance of the average AGB star, 
it is not surprising that such small scale structure usually is unresolved in
other systems.  Small mid-IR disks around AGB stars could be 
common but difficult to resolve.  

We model the IR emission of RV Boo as an optically thin disk
passively heated by the star.\footnote{%
  RV Boo has a $B$--$V$ color of $\simali$1.47\,mag;
  this low reddening implies that the mid-IR emitting
  region is not obscuring the entire star (as it would
  be if the dust was spherically distributed around the star)
  and is largely concentrated along one plane. 
  Thus, we have explicitly modeled the IR emission
  around RV Boo as a disk. Since we have {\it apriori} 
  considered a disk morphology, our model cannot
  constrain the vertical geometry of the emission.
  } 
We approximate the stellar 
photosphere by the \citet{Kurucz} model spectrum for M6IIIe 
stars with an effective temperature of $\Teff=3000\K$.
The dust is taken to be amorphous silicate since RV Boo
is an M star. We assume a power-law dust size distribution 
$dn(a)/da \propto a^{-\alpha}$ which is characterized by 
a lower-cutoff $\amin$, upper-cutoff $\amax$ 
and power-law index $\alpha$.\footnote{%
  We assume all grains are spherical in shape
  with $a$ being their spherical radius.
  }
We take $\amin=0.01\mum$ since grains smaller than this will
undergo single-photon heating \citep{Draine2}
while the observed IR spectral energy distribution
of RV Boo does not appear to show evidence for 
stochastically heated dust, and $\amax=1000\mum$ since 
larger grains are not well constrained by 
the currently available IR photometry.

The dust spatial distribution is taken to be 
a modified power-law
$dn/dr \propto \left(1-\rmin/r\right)^{\beta} 
\left(\rmin/r\right)^{\gamma}$
where $\rmin$ is the inner boundary of the disk
which we take to be the location where silicate
dust sublimates. For RV Boo with a luminosity
$\Lstar\approx 8100\,\Lsun$ \citep{RVBoo1}
submicron-sized silicate dust achieves 
an equilibrium temperature of $T\approx 1500\K$ 
and starts to sublimate at $r\approx 3\AU$. 
Therefore we take $\rmin=3\AU$. 
This functional form has the advantage that on one hand,
it behaves like a power-law $dn/dr \propto r^{-\gamma}$ at
larger distances ($r\gg\rmin$), and on the other hand
it peaks at $\rp = \rmin \left(\beta+\gamma\right)/\gamma$,
unlike the simple power-law which peaks at $\rmin$. 
The latter is unphysical since one should not expect 
dust to pile up at $\rmin$ where dust sublimates! 
We take $\gamma=2$ as expected from a stationary outflow.
We take $\rmax=120\AU$ which is large enough for our dust
IR emission modeling purpose since there is very little dust
beyond this outer boundary.  
Therefore, we have only two free parameters: 
$\alpha$ -- the power-law exponent for the dust size
distribution and $\beta$ -- the dust spatial distribution
parameter which determines where the dust peaks.

\begin{figure}
\epsscale{0.6}
\includegraphics[width=\columnwidth]{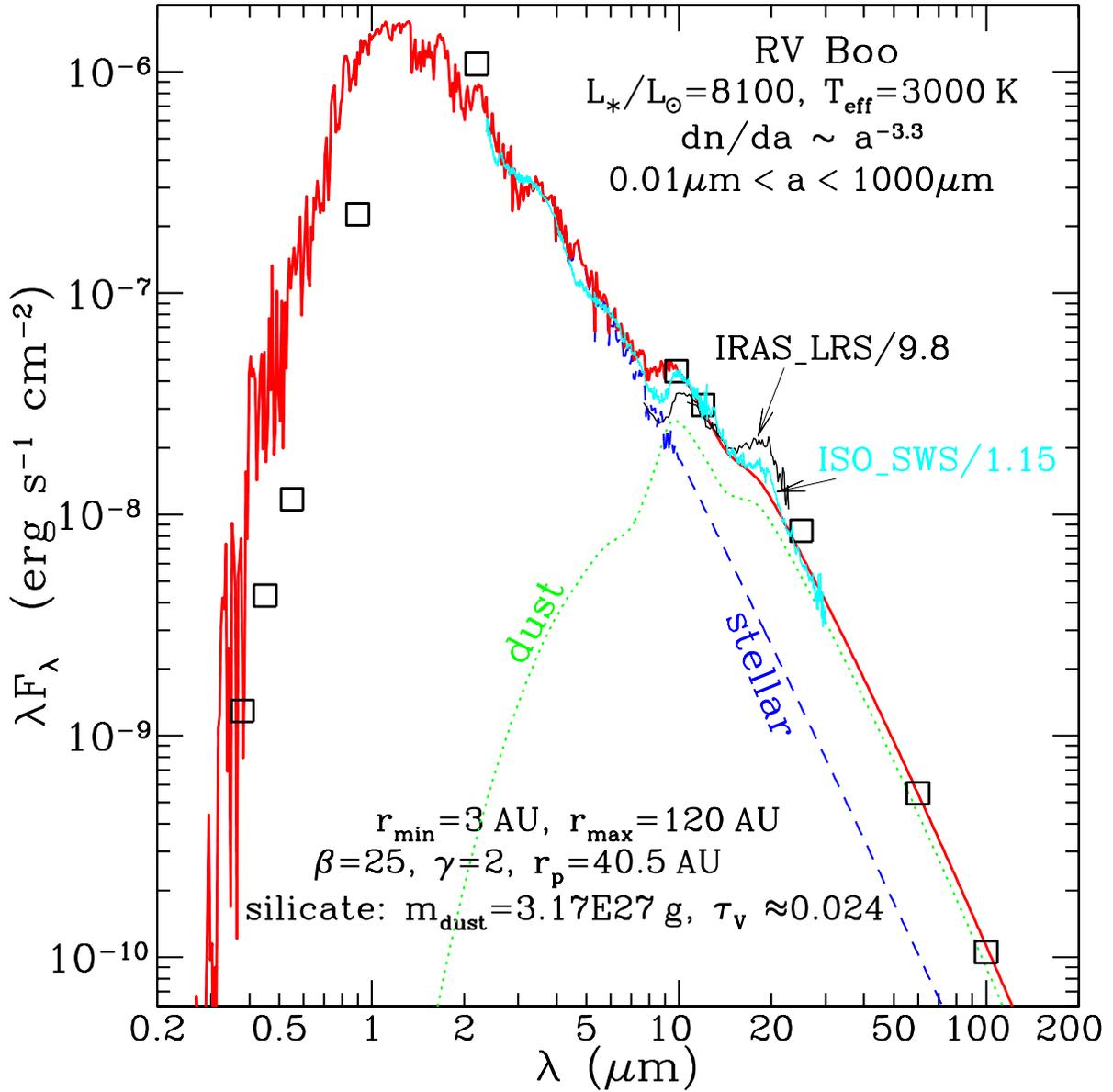}
\caption{
IR emission and model fit to the RV Boo disk.
        The open black squares are the U, B, V, I, K,
        the 9.8$\mum$ AO photometry,
        and the 12, 25, 60 and 100$\mum$ IRAS photometry.
        Thin solid black line is the IRAS LRS spectrum
        (reduced by a factor of 9.8 in order to agree with
         the IRAS 12$\mum$ data; see \S4).
        Thin solid cyan line is the ISO SWS spectrum
        (reduced by a factor of 1.15; see \S4;
         the spectrum longward of 30$\mum$ is too noisy
         and therefore not shown in this figure).
        Green dotted line is the dust model emission.
        Blue dashed line is the stellar photospheric spectrum
        approximated by the Kurucz model of $\Teff=3000\K$.
        Red solid line is the sum of the dust and stellar emission.
        At a distance of $d\approx 390\pc$, this best fit model
        has an inner edge at 3$\AU$ and a peak dust density at
        $r_{\rm p} \approx 40.5\AU$.
}
\label{fig:RVBoo_sed}
\end{figure}

\begin{figure}
\plotone{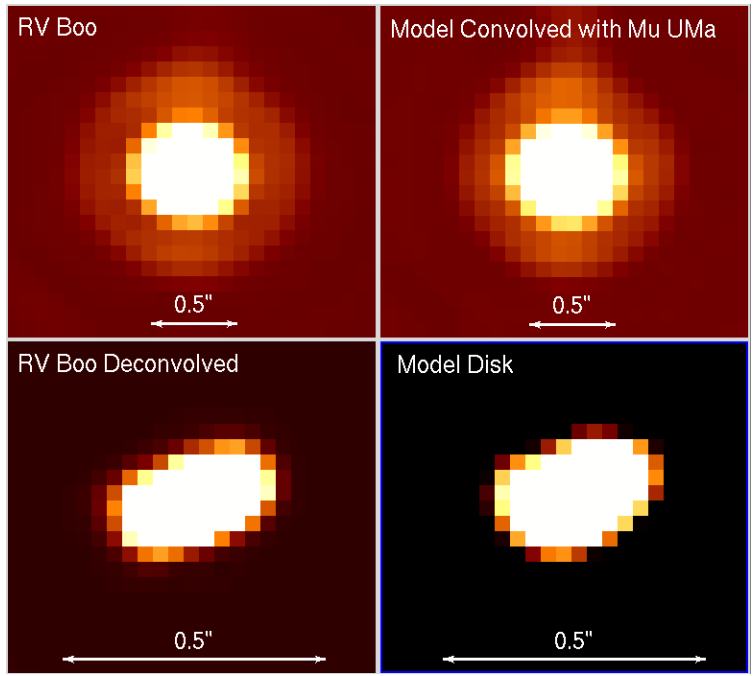}
\caption{Comparison of RV Boo to the best fit dust disk model 
(r$_{inner}$ = 3 AU, r$_p$ = 40 AU, $\alpha\approx 3.3$, $\beta\approx 25$,
total dust mass $m_{\rm dust}\approx 3.17\times 10^{27}\g
\approx 1.6\times 10^{-6} M_{\odot}$ and an inclination from 
edge on of 40$^{\circ}$.)  
For comparison with the raw RV Boo image, the model image in the lower
right has been convolved with the $\mu$ UMa PSF.  North is up and east is to 
the left in all these images.  Note that 
the RV Boo data is consistent with 
the best fit SED model from Fig.~\ref{fig:RVBoo_sed}.}
\label{fig:RVBoo_mod}
\end{figure}

Using the dielectric functions of \citet{Draine1}
for ``astronomical silicates'' and Mie theory \citep{Bohren}, we calculate
the absorption cross sections of amorphous silicate grains
as a function of size and their steady-state temperatures
(as a function of radial distance from the central star)
in thermal equilibrium with the illuminating starlight
intensity. We then obtain the dust model IR emission
spectrum by integrating the dust emission over the entire
size range and the entire disk and compare with
the 12, 25, 60, and 100$\mum$ IRAS ({\it Infrared Astronomical
Satellite}) broadband photometry,
the 7.7--22.7$\mum$ low resolution spectrum
obtained with the IRAS {\it Low Resolution Spectrometer}
(LRS; with a resolution of $\lambda/\Delta\lambda \approx 20$),
the 2.5--45$\mum$ high resolution spectrum obtained with
the {\it Short-Wavelength Spectrometer} (SWS) instrument
(with a resolution of $\lambda/\Delta\lambda \approx 2000$)
on board the {\it Infrared Space Observatory},\footnote{%
  The IRAS LRS spectrum integrated with the IRAS 12$\mum$
  band filter function results in a factor of $\sim 9.8$
  higher than the IRAS 12$\mum$ photometric flux.
  This is not surprising in view of the fact
  that RV Boo is a variable star.
  We have therefore reduced the flux level of the IRAS LRS spectrum
  by a factor of 9.8 to bring it into agreement
  with the IRAS 12$\mum$ broadband photometric data.
  Similarly, the flux level of the ISO SWS spectrum
  has also been reduced by a factor of $\sim 1.15$,
  in order to agree with the IRAS 12$\mum$ photometry.
  }
and the 9.8$\mum$ photometry presented
in this work. We also compare the dust model image at 9.8$\mum$
convolved with the AO instrument PSF with the AO image of
RV Boo and inclined at a variety of inclinations from edge on.  A
number of models with $\beta\approx 25-30$ and inclinations from
edge on of 30-45$^{\circ}$ fit the data to within 20$\%$ accuracy.
In Figures~\ref{fig:RVBoo_sed} and~\ref{fig:RVBoo_mod},
we respectively show the best-fit IR emission spectrum and disk
image.  The disk image is also shown convolved with the instrument PSF.
spectrum (see Fig.~\ref{fig:RVBoo_sed}) and the AO image
(see Fig.~\ref{fig:RVBoo_mod}).
The source properties (eccentricity, FWHM)
for this model are plotted alongside source properties for the telescope data
and PSF stars in Fig.~\ref{fig:E}.
This model, with $\alpha\approx 3.3$ and $\beta\approx 25$,
has a total dust mass $m_{\rm dust}\approx 3.17\times 10^{27}\g
\approx 1.6\times 10^{-6} M_{\odot}$ and an inclination from
edge on of 40$^{\circ}$. The dust spatial distribution peaks
at $\rp\approx 40.5\AU$. The maximum vertical visual optical
depth is $\tau_V\approx 0.024$, confirming the validity of
the optical-thin assumption made at the beginning
of this modeling effort.

\section{Discussion \label{sec:discussion}}

Where do the wide variety of PNe shapes come from?  Mass loss has 
practically ended by the PPN stage; therefore, any underlying structure which
produces the shape of PN (via theories such as the Generalized
Interacting Stellar Wind models) must form by the end of the 
AGB stage.  
While PNe possess an extremely wide range of morphologies, \citet{PPNsurvey} 
found two primary mid-IR morphologies for PPNe 
corresponding to different values of 
optical depth -- optically thick core/elliptical morphologies and 
optically thin toroidal morphologies.  \citet{PPNsnapshot}
found that each of these 
mid-IR morphologies also corresponds to a specific optical morphology.  
Optically thick
core/elliptical mid-IR morphologies possess bipolar reflection nebulae
and heavily obscured central stars (called DUPLEX by \citet{PPNsnapshot})
while optically thin toroidal mid-IR
morphologies possess elliptical reflection nebulae and non-obscured
central stars (called SOLE by \citet{PPNsnapshot}).

The PPNe observed by \citet{PPNsurvey} are more evolved objects than AGB stars.
Since mass loss has ceased by the PPN stage, the dense core/elliptical 
dust structures which produce DUPLEX sources (and are the precursors
to structures formed during the PN stage) must have formed during the 
end of the AGB phases.  AGB mass loss can be divided into two separate phases:
an initial, spherically symmetric AGB wind ($\sim$10 km s$^{-1}$), 
supplanted by 
a faster superwind ($\sim$20 km s$^{-1}$) for a brief period at the end of the 
AGB phase (\citet{PPNsnapshot}, \citet{superwind}).  
With the onset of the superwind,
mass loss rates are expected to rise by factors of $\geq$10 \citep{Steffen}.
\citet{PPNsnapshot} propose that this 
superwind at the end of the AGB phases 
is intrinsically asymmetric, producing an equatorially flattened toroid which 
collimates later bipolar structure.  However, it is not clear what mechanism
could produce the asymmetry.

An AGB star like RV Boo which displays asymmetric structure 
is an excellent laboratory 
for studying the very beginnings of PN and PPN structure formation. 
RV Boo is one member of a set of AGB stars which display hallmarks of 
asymmetric structure.  This small group of AGB stars ($\sim$20$\%$) display 
very narrow CO linewidths 
(as narrow as $\sim$1 km s$^{-1}$, see Kahane et al. 1998, but 
generally $\leq$5 km s$^{-1}$), with 
or without a broader underlying pedestal feature (with widths of 10-20 km/s).  
BM Gem displays just a narrow peak, while
X Her \citep{XHer}, RV Boo \citep{RVBoo1}, EP Aqr, RS Cnc, and
IRC +50049 display narrow peaks as well as an underlying broader component. 
\citet{MolRes} interpret the narrow features in these objects as reservoirs 
of dust and molecular gas which are nearly at rest with 
respect to the central AGB stars.  
The broader CO line components are interpreted as spherical outflows
or in some cases (RS Cnc and X Her) as bipolar outflows \citep{XHer}.
  
RV Boo is unusual among these stars since images of its CO emission suggest 
the presence of a large disk in Keplerian rotation \citep{RVBoo1}.  One 
possible explanation for the presence of the disk 
is that mass loss from the AGB star has become
equatorially enhanced through entrainment by a binary companion 
\citep{MolRes, binary, Mas1, Mas2}.
Could this be the mechanism (in progress) which 
produces the dense collimating toroid of material invoked by 
\citet{PPNsnapshot} -- essentially, Ueta's equatorial superwind?  While
RV Boo does not have a known companion, a close companion ($>$100 AU) may be 
currently undetectable.

Interestingly, similar narrow CO line structure has been observed around a
number of evolved post-AGB systems -- two RV Tauri stars, AC Her (which has
only a narrow line without a broad pedestal, suggesting perhaps 
that mass loss has ceased and only a molecular reservoir remains) and 
IRAS 08544-4431 \citep{IRAS08544} and the PPN Red Rectangle \citep{RedRect1}.
Additionally, a dusty disk has been detected around IRAS 08544-4431  
and a Keplerian disk has been detected in CO around the PPN 
Red Rectangle \citep{RedRectDisk}.  Are the disks around IRAS 08544-4431 and 
the Red Rectangle similar to that around RV Boo, but at the corresponding 
later stages of evolution?  All three of these objects have 
known companions, a fact which supports 
theories of binary entrainment for disk formation.
               
If RV Boo is nearing the end of its AGB phase, the disk structure observed 
in CO and in the mid-IR may be the beginning of a denser disk/torus
capable of collimating the fast winds produced during the PN phase.
However, to determine whether the structure around RV Boo is the precursor of 
PPN and PN structure requires knowledge of the evolutionary status of RV Boo 
and also whether objects of a similar evolutionary status possess similar
CO and mid-IR structure.
RV Boo is a ``red'' SRb, according to the classification system of 
\citet{SRV1}.  They interpret the red SRb's as a transitional stage
between the blue SRb's, which sit near the tip of the Red Giant branch, and 
Miras, which mark the tail end of the AGB phase.  With a moderate mass
loss rate of $\sim$10$^{-7}$ M$_{\odot}$ yr$^{-1}$, RV Boo is probably not at 
the very tail end of AGB evolution.  However, \citet{SRV1} use single-star 
stellar models to constrain evolutionary state. 
If disk formation does require a binary companion, estimates of evolutionary
state from such models may be inaccurate for RV Boo and similar objects. 

While narrow CO lines have been found in $\sim$20$\%$ of AGB stars, it
remains to be seen how common extended 
mid-IR structure such as that observed around RV Boo occurs in AGB stars.  
Indeed, even if mid-IR structure around narrow CO line AGB stars is common, 
it remains to be seen how similar RV Boo's CO and mid-IR structures
are to those of other AGB stars with narrow CO lines.
This is a comparison we can not yet make, 
since no other similar 
object has both been mapped interferometrically in CO and imaged at high
resolution in the mid-IR.
Several of these objects 
share very similar observational properties with RV Boo -- in particular,
X Her, EP Aqr, RS CNc, BM Gem, and EU And.  It would be worthwhile 
observe these objects both at high resolution in the mid-IR and also 
interferometrically in CO.

\section{Conclusions \label{sec:conc}}

We present the first high resolution (0.1\arcsec) very high 
Strehl ratio (0.97$\pm$0.03) mid-IR images of 
RV Boo utilizing the MMT adaptive secondary
AO system. These are the first IR images presented in the literature with
such high Strehl ratios; previous  
Strehl ratios for large telescopes have hardly exceeded 70$\%$ at any 
wavelength.  RV Boo was observed at a number of wavelengths over two epochs 
(9.8 $\mu$m in May 2003, 8.8, 9.8 and 11.7 $\mu$m in February 2004)
and appeared slightly extended at all wavelengths.  
While the extension is within the 
measurement error for the 8.8 and 11.7 $\mu$m data, 
the extension is more pronounced in the 9.8 $\mu$m data.
The slight extension seen in the 9.8 $mu$m data from both May 2003 and 
February 2004 suggests that the mid-infrared structure around 
RV Boo is marginally resolved at 9.8 $\mu$m.  

Because of our high Strehl ratios which 
leads to extremely stable PSFs, we can deconvolve
our images with those of PSF stars for a super-resolution of 0.1\arcsec.
Based on the rotation of the extension with the sky 
and its non-PSF FWHM (at 8.8-11.7 $\mu$m) and eccentricity, 
we conclude that the extension around RV Boo
is indeed real and not an artifact of the telescope.  
 
We tentatively detect
 a $\sim$60 AU FWHM (0.16$\arcsec$ at 390 pc) disk at 9.8 $\mu$m around
RV Boo.  Previously, \citet{RVBoo1} found a 4$\arcsec$ disk in CO with a 
position angle of 150$^{\circ}$; 
we find a position angle of 120$\pm$6$^{\circ}$ for the 9.8 $\mu$m 
disk.   We measure a total disk flux of 145$\pm$24 Jy at 9.8 $\mu$m.
We closely reproduce the observed IR spectral energy distribution 
and the AO image in terms of an optically thin dust disk consisting of 
amorphous silicates with a power-law size distribution. 
We estimate a disk inclination
angle of 30 to 45$^{\circ}$ from edge on and a 
disk dust mass of 1.6$\times$10$^{-6}$ M$_{\sun}$.

\acknowledgements
These MMT observations were made possible by the hard work of the 
entire Center for Astronomical Adaptive Optics (CAAO) staff 
at the University of 
Arizona.  In particular, we would like to thank Tom McMahon, Kim Chapman, Doris
Tucker, and Sherry Weber for their endless support of this project.  The wide
field AO CCD was installed by graduate student Nick Siegler.  Graduate student 
Wilson Liu helped run the MIRAC3 camera during the run.  The adaptive secondary
mirror is a joint project of University of Arizona and the Italian National 
Institute of Astrophysics -- Arcetri Observatory.  We would also like to thank the
whole MMT staff for their excellent support and flexibility during our 
commissioning run at the telescope.

The secondary mirror development was supported by the Air Force 
Office of Scientific Research under grant 
AFOSR F49620-00-1-0294.  BAB, LMC, and DP acknowledge 
support from NASA Origins 
grant NAG5-12086 and NSF SAA grant AST0206351.  JHB acknowledges support from
NSF grants AST-9987408 and AST-0307687.

\end{document}